\documentclass[12pt,letterpaper]{article}
\usepackage[T1]{fontenc}
\usepackage[utf8]{inputenc}
\usepackage{mathptmx}
\usepackage[margin=1in]{geometry}
\usepackage{graphicx}
\usepackage{amsmath,amssymb}
\usepackage{cite}
\usepackage{titlesec}
\usepackage[font=small,labelfont=it,textfont=it,justification=justified,singlelinecheck=false]{caption}
\usepackage[hidelinks]{hyperref}
\usepackage{url}

\titleformat{\section}{\large\bfseries}{\thesection.}{0.6em}{}
\titleformat{\subsection}{\normalsize\bfseries}{\thesubsection}{0.6em}{}
\titlespacing*{\section}{0pt}{16pt}{6pt}
\titlespacing*{\subsection}{0pt}{12pt}{4pt}
\renewcommand{\thesubsection}{\thesection.\arabic{subsection}}
\begin{document}

\begin{center}
{\LARGE\bfseries An AI Factory for Technology Education: An Agentic Pipeline with Closed Loop Audience Feedback\par}
\vspace{14pt}
\begin{tabular}[t]{@{}c@{}}
\textbf{Josue Hernandez}\\
Dapta Inc., Phoenix, USA\\
\texttt{josue@daptatech.com}\\
ORCID: 0009-0005-5323-1405
\end{tabular}

\vspace{8pt}
\begin{tabular}[t]{@{}c@{}}
\textbf{Hector R. Amado-Salvatierra}\\
Universidad Galileo, Guatemala, Guatemala\\
\texttt{hr\_amado@galileo.edu}\\
ORCID: 0000-0002-3028-234X
\end{tabular}

\vspace{8pt}
\begin{tabular}[t]{@{}c@{}}
\textbf{Rocael Hernandez Rizzardini}\\
Universidad Galileo, Guatemala, Guatemala\\
\texttt{roc@galileo.edu}\\
ORCID: 0009-0000-8818-5617
\end{tabular}
\end{center}

\vspace{6pt}
\noindent\textbf{Abstract}

\begingroup\setlength{\parskip}{0pt}
Technology tools used for professional upskilling change on a timescale of weeks, and written guidance about them ages at the same pace. Large language models have made it inexpensive to draft a tutorial or a course outline, but most deployed generation systems still produce a piece of content once and never check it against what an audience actually does with it. This paper describes an agentic pipeline, organized as a typed workflow graph, that treats instructional content production as a manufacturing process with explicit checkpoints rather than a single generation call. Four node classes structure the graph: deterministic scripts, model steps bound to a fixed rubric, open-ended agent phases, and human approval gates, arranged along a forward production path and closed by a feedback cycle that rewrites generation rules from post level engagement telemetry. Production of content is split into two layers: a demand layer, built from video and community sources, that shapes topic selection and register, and an authority layer, built from official documentation, that every published claim must trace back to. Over five months of continuous operation, the deployed system published 26 documented guides at roughly 1.5 hours of operator time per week, and the feedback loop identified four measurable engagement effects that became binding rules for later runs, including a hook style that outperformed baseline by a factor of 1.83 in its first analysis window.
\par\endgroup


\section{Introduction}\label{sec:intro}

Software tools, application programming interfaces, and model releases in the artificial intelligence space change often enough that a written guide can lose accuracy within months of publication. Producing that guidance at the pace the tools themselves change is largely a content generation problem, and large language models have lowered the cost of drafting a tutorial, a comparison of two competing tools, or a full course outline from a syllabus or a topic prompt. A growing body of work has used language models for this exact purpose, from single prompts that draft a course~\cite{rouabhia2024,dickey2024,macneil2023} to one or several agents that research a topic, draft instructional text, and in some cases assemble a full lesson plan~\cite{yao2026,zhang2025}.

However, most of these systems are built around a single generation pass. A topic goes in, an agent or a short agent chain produces a document, and the document is either published as is or handed to a human editor with no further mechanism to check it against external sources or to learn from how an audience actually responds to it. Two properties of this design are hard to see from a single deployment but compound over a production run of dozens of guides. First, the boundary between steps that require judgment and steps that are purely mechanical is rarely made explicit, so the same language model call ends up doing topic scoring, drafting, and formatting checks with no differentiated cost or fixed rubric. Second, there is commonly no path back into the system: engagement with published content, when it is measured at all, does not change what the pipeline does next.

This paper presents an agentic pipeline, deployed in continuous production for content on artificial intelligence tools, that addresses both properties directly. The pipeline is described as a typed workflow graph, a directed graph whose nodes fall into four classes by the kind of work they perform: deterministic scripts with fixed behavior, model steps that apply a fixed scoring rubric, open ended agent phases that carry out research and writing, and gate nodes where a human operator approves or sends work back for revision. Content flows forward through two research phases, a generation phase, and publication; a second, dashed cycle runs across production runs rather than within one, carrying post level engagement telemetry back into an autonomous analysis step that rewrites a shared strategy file read by both topic selection and generation in every subsequent run.

Framing the pipeline this way, as a typed graph with an explicit feedback cycle, has two consequences developed later in this paper. Structurally, it lets deterministic, model bound, and open ended work be assigned to the cheapest node class capable of it, reserving full agent invocations for the three phases that actually require open ended judgment. Empirically, it lets the system be evaluated on production evidence rather than on a single generated sample: over five months of operation the pipeline produced 26 documented multi-page guides, and the feedback cycle logged four measurable engagement effects, one of which reversed an earlier default once later data contradicted it.

The remainder of this paper is organized as follows. Section~\ref{sec:related} reviews related work on agentic content generation systems, agent harness design, and closed loop learning analytics. Section~\ref{sec:arch} describes the system architecture in full, covering the orchestrator and executor split, the four node classes, and each of the three production phases together with the feedback cycle that closes across them. Section~\ref{sec:design} presents the content design model implemented inside the generation phase, including the artifact inventory, the content taxonomy, and the embedded instructional scaffolding. Section~\ref{sec:results} reports production data collected over five months of deployment, including operator effort, the instrumented acquisition funnel, and the specific findings the feedback loop has produced so far. Section~\ref{sec:discussion} discusses the design choices behind the architecture, Section~\ref{sec:limits} states the limitations of the current evidence and the planned extensions, and Section~\ref{sec:conclusion} concludes the paper.

\section{Related Work}\label{sec:related}

Language model systems for generating instructional content range from single prompt tools to coordinated multi agent systems. Rouabhia reports a case study using ChatGPT to draft a full course from a syllabus~\cite{rouabhia2024}, and Dickey and Bejarano propose a framework, GAIDE, for using generative AI to assist course content development~\cite{dickey2024}. MacNeil et al.~show that large language models can automatically produce learning materials for a computer science course at a quality comparable to instructor written material~\cite{macneil2023}. In the MOOC setting, Amado-Salvatierra et al.~examine the benefits and limitations of combining human creativity with AI-based tools in instructional design~\cite{amado2023}, and Rizzardini et al.~integrate a conversational pedagogical agent into the instructional activities of a MOOC~\cite{rizzardini2021}. Beyond a single generation call, several systems assign separate agents to separate stages of the task: Instructional Agents splits lesson design across a small team of agents to reduce faculty workload~\cite{yao2026}, EduPlanner applies a similar multi-agent structure to customized instructional design~\cite{zhang2025}, and MAIC reshapes online teaching and learning through a set of coordinated language model driven agents~\cite{yu2024}. AgentLesson applies a multi-agent architecture specifically to lesson plan generation~\cite{chen2026}, and a comparable pattern recurs beyond text: EduIllustrate generates multimodal educational content at scale~\cite{bi2026}, an LLM based multi-agent system produces full educational videos from a script~\cite{yan2026}, and a multi-agent workflow evaluates procedural learning content for pedagogic quality at scale~\cite{pei2025}. Retrieval augmented generation has also been applied to specific artifact types adjacent to guide production, including quiz generation~\cite{gopi2024} and automated test generation and marking~\cite{papachristou2025}.

A parallel line of work treats the infrastructure around a language model, rather than the model itself, as the object of study. Recent surveys describe this infrastructure as an agent harness: the orchestration layer that assembles context, exposes tools, sequences turns, and carries governance across a task that spans multiple steps~\cite{meng2026,ali2026,chen2024survey,li2024survey,zhu2026}. Pathak frames the same idea in practitioner terms, arguing that the harness, not the underlying model, determines whether the same model behaves reliably in production~\cite{pathak2026}. Ning et al.~propose treating code itself as the harness for agent behavior~\cite{ning2026}, AutoGen defines a general framework for coordinating conversations among multiple agents~\cite{wu2023}, and Jiang et al.~survey agentic workflow concepts and applications specifically for education~\cite{jiang2025}. Kamalov et al.~review the broader evolution of agentic workflows applied to educational tasks~\cite{kamalov2025}.

A separate strand of research closes the loop between learner or audience behavior and system adaptation, mostly within learning analytics rather than content generation. Clow describes the learning analytics cycle as a loop that must close from data back to intervention to be effective~\cite{clow2012}, and Sailer et al.~formalize a similar closed loop learning analytics framework~\cite{sailer2024}. Syed et al.~implement an integrated closed loop learning analytics scheme in a first year experience course~\cite{syed2019}, Ninaus and Sailer examine the human role that should remain inside such a loop~\cite{ninaus2022}, and Wang et al.~combine neural cognitive diagnosis, adaptive testing, and language model driven feedback into a closed loop personalized learning agent~\cite{wang2025}. Khalil et al.~use learning analytics to iteratively improve the educational design of MOOCs~\cite{khalil2022}, and Uzun et al.~study how higher education students interact with a generative AI assistant answering questions about their own learning analytics feedback~\cite{uzun2026}. Tarun et al.~propose human in the loop systems for adaptive learning with generative AI, keeping a human reviewer inside the adaptation cycle itself~\cite{tarun2025}.

The pipeline presented here combines three properties: a typed workflow graph that separates deterministic, model bound, and open ended work into distinct node classes, each carrying its own cost and rubric; two layer source grounding that keeps audience demand and factual authority structurally separate; and a feedback cycle that rewrites its own generation rules from engagement telemetry while a human operator retains override authority over every rule it produces. However, no system reviewed above combines all three. The multi-agent content generation systems described above assign specialized agents to specialized tasks, but none differentiates that work into a typed graph with a distinct cost and rubric per node class. The agent harness literature analyzes orchestration as an economic and reliability problem, but largely for general purpose agentic applications rather than for a concrete content production pipeline with human approval gates. A related strand of research closes a similar loop from learner data back to an intervention, though that intervention is typically an assessment or a recommendation, not the rules that generate the content itself. This gap motivates the pipeline presented in this paper.

\section{System Architecture}\label{sec:arch}

\begin{figure}[p]
\centering
\includegraphics[width=0.97\textwidth]{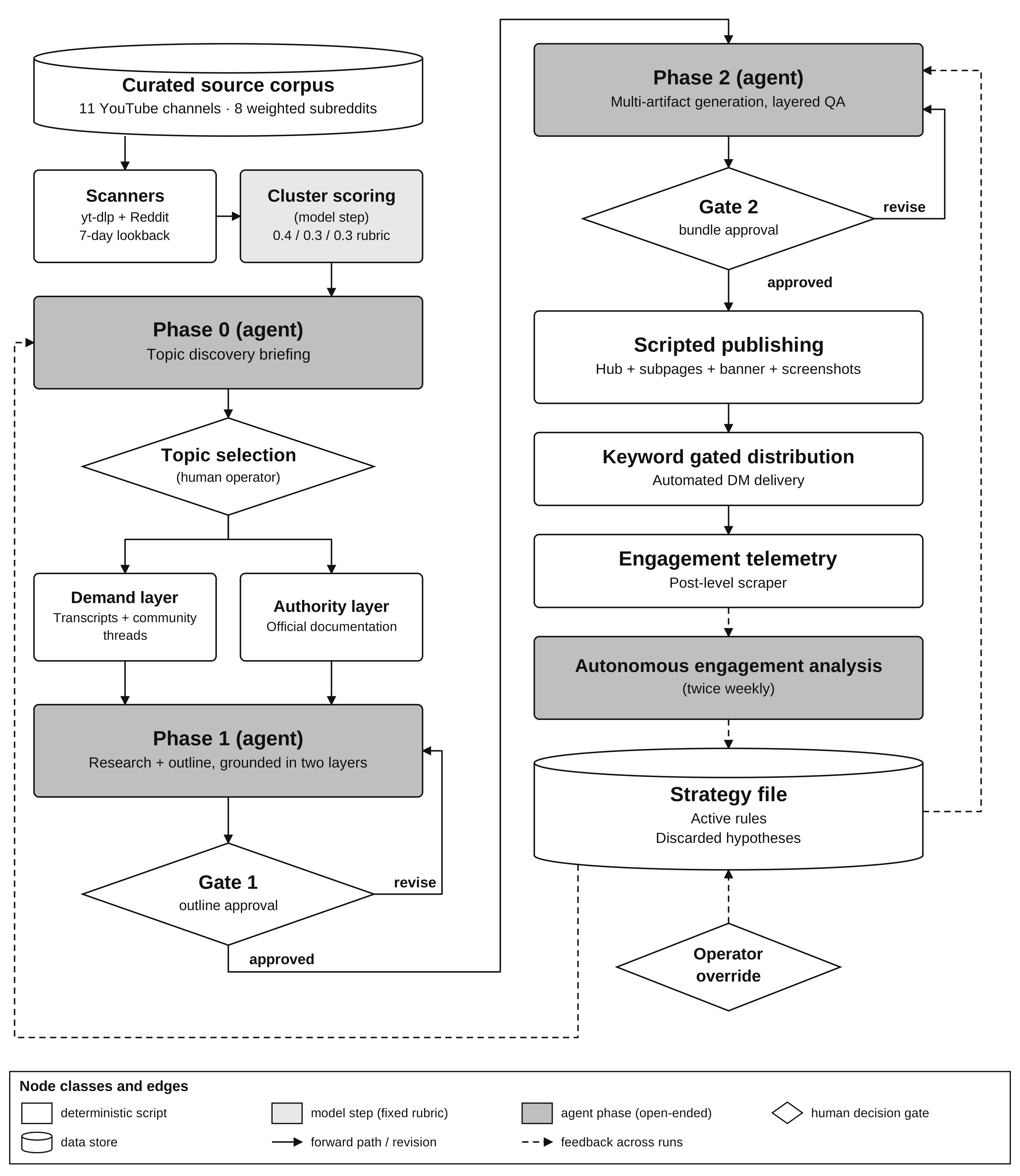}
\caption{The complete workflow graph, drawn top to bottom. Node shape and fill encode class: white rectangles are deterministic scripts, light gray rectangles are model steps bound to a fixed rubric, dark gray rectangles are open ended agent phases, diamonds are human decision gates, and cylinders are data stores. Solid edges form the forward production path from the curated source corpus to keyword gated distribution. Dashed edges form the feedback cycle that runs across production runs: engagement telemetry feeds an autonomous analysis step that rewrites the strategy file, which both Phase 0 and Phase 2 read on every subsequent run; the operator can override any rule the loop produces, and the loop adapts to that override in its next cycle.}
\label{fig:workflow}
\end{figure}

This section presents the pipeline at the level of detail needed to reimplement it. An orchestrator process handles every step that touches the operator or an external system, while a single executor agent carries out the three open ended production phases: topic research, source grounded outlining, and multi artifact writing. Four node classes compose the resulting workflow graph, distinguished by how much judgment each step requires: a script carries none, a model step follows a fixed rubric, an agent phase reasons openly, and a gate hands the decision to the operator. A fifth mechanism closes a feedback cycle across runs: engagement telemetry rewrites a strategy file that later runs read, subject to the operator\textquotesingle s override authority. Figure~\ref{fig:workflow} shows the complete graph top to bottom.

\subsection{Orchestrator and Executor Decomposition}

The system separates interactive coordination from autonomous production. An orchestrator process, defined in a skill specification, owns every step that requires human contact or a side effect on an external system: presenting a briefing or an outline for review, publishing to the hosting platform, generating cover graphics, and creating distribution records. A single executor agent, defined in a separate agent specification, performs the phases that require open ended judgment: topic research, source research and outlining, and full content generation. The orchestrator spawns the executor once per phase, execution is strictly sequential with no parallel branching and no dedicated reviewer agent, and quality control is embedded in the executor itself as mandatory checklists rather than delegated to a second model call.

The executor carries persistent memory scoped to the project: a directory whose index file loads into the agent's system prompt at every invocation and survives across sessions. The agent is instructed to record which guide structures were approved versus heavily edited, which hook formulas performed well, which topics or angles were rejected, and which issues kept getting caught during human review. This turns per run editorial correction into standing knowledge the system carries forward, so a correction is paid for once rather than repeated on every run.

\subsection{The Workflow as a Typed Graph}

The pipeline is best described as a directed graph whose nodes fall into four classes. Deterministic nodes are scripts with fixed behavior: channel and forum scanners, publishing scripts, and a graphic generator that produces banners and distribution images conditioned on reference logo images. Model nodes perform a bounded judgment task against a fixed rubric, such as classifying content type or scoring a topic cluster. Agent nodes are the executor's open ended phases: research, outlining, and multi-artifact writing. Gate nodes are the human decision points: topic selection from the Phase 0 briefing, plus the two formal approval points, where the operator approves the outline after Phase 1 and the full content bundle after Phase 2. This four-class split is an analytical decomposition by the kind of work each step performs, not by process boundary; classification and scoring run inside the executor's phases rather than as separate agent invocations. The rules governing the workflow limit interruption between the two gates deliberately: once the outline is approved, generation proceeds on named defaults, such as a fixed subpage count and a specified lead hook variant, and the executor may ask at most one consolidated question rather than a series of them.

The distinguishing edge in this graph sits not on the forward path but across production runs. An autonomous engagement analysis step, external to the main pipeline, rewrites a strategy file from observed post level engagement telemetry, and that file is a required input to both topic discovery and content generation. Content flows along the forward path; adaptation flows along this second cycle, and it is the presence of this cycle, closing rules back into the pipeline instead of only reporting on past posts, that distinguishes the system from an ordinary linear content pipeline.

\begin{figure}[p]
\centering
\includegraphics[width=0.88\textwidth]{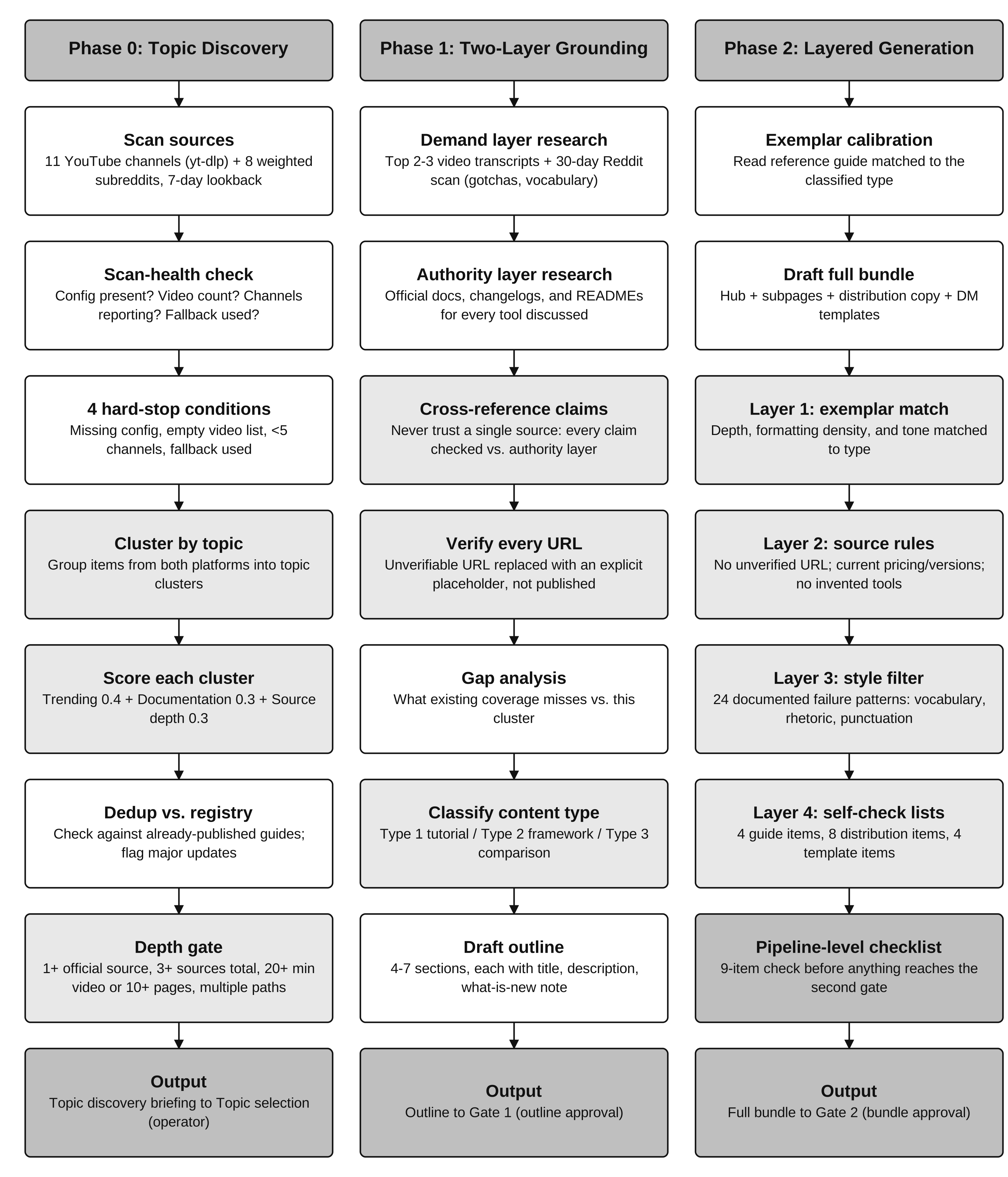}
\caption{Internal steps of Phase 0, Phase 1, and Phase 2. Fill follows the same node classes as Figure~\ref{fig:workflow}: white steps are deterministic or purely procedural, light gray steps apply a fixed rubric or rule, and dark gray steps are the phase boundary itself or the pipeline-level checklist that gates the return to the orchestrator.}
\label{fig:phases}
\end{figure}

Figure~\ref{fig:phases} unpacks the internal steps of each of the three production phases, side by side.

\subsection{Phase 0: Topic Discovery}

Topic discovery runs on demand against a curated source corpus: 11 configured YouTube channels (the configuration in force during the March to July 2026 deployment reported in Section~\ref{sec:results}), scanned through yt-dlp over a 7 day lookback window, and 8 subreddits, scanned with per-subreddit weights so that signals from the community most relevant to the topic domain rank above general automation forums. Items from both platforms are clustered by topic, and each cluster is scored on three dimensions with fixed weights: trending signal at 0.4, documentation availability at 0.3, and source depth at 0.3, combined linearly into a single composite score. A topic appearing on both platforms in the same week counts as the strongest available demand signal; within a single platform, coverage by three or more channels, or a single post with 200 or more upvotes, marks a strong cluster.

Two filters run after scoring. First, candidate topics are checked against the registry of already published guides, and major updates to a covered topic are flagged separately as update opportunities rather than silently re-ranked alongside new topics. Second, a depth gate rejects any topic that cannot sustain a full guide regardless of how strongly it trends. The gate requires at least one authoritative official source, three or more sources in total, either 20 or more minutes of video or 10 or more pages of written material, more than one valid implementation path, and enough substance to fill four to seven sections each containing real code, commands, or configuration. News items, single feature announcements without documentation, and pure opinion content are rejected outright.

Phase 0 is designed to fail loudly rather than degrade quietly. Every briefing opens with a scan health block reporting whether the channel configuration was present, how many videos the scanner returned and across how many channels, the lookback cutoff applied, and whether a web search fallback was used. Four conditions force a hard stop instead of a ranking: a missing channel configuration, an empty video list from the scanner, fewer than 5 channels returning results, or a scanner exception followed by a web search fallback. The specification treats a silent fallback as the worst possible outcome, citing a documented incident on a specific date in which a fallback built briefing missed a directly relevant topic while still appearing authoritative.

\subsection{Phase 1: Grounding in Two Layers}

Research separates a demand layer from an authority layer. The demand layer establishes what the audience wants and how it talks about it: yt-dlp extracts auto generated transcripts from the top two or three videos in the selected cluster, and a 30 day Reddit scan mines community threads for unexplained problems, debated alternatives, real audience complaints, and the vocabulary needed to calibrate the writing voice. The authority layer establishes what is actually true: official documentation, changelogs, and repository readme files are fetched for every tool discussed, and every claim a video creator makes is checked against these primary sources under an explicit rule never to trust a single source.

An obvious objection is that a pipeline seeded by social video content inherits the reliability problems of social video. The answer built into the design is structural rather than a matter of intent: social sources are confined to the demand layer. They determine topic selection, coverage priorities, and register, but no factual claim is published on their word alone. Any URL that cannot be verified is replaced with an explicit placeholder rather than published, the source list attached to every guide is restricted to evergreen authoritative material, and video sources may never appear in that published list. A fact learned only from a video must be derived from an official source and cited there instead. Phase 1 closes with a gap analysis identifying what existing coverage misses, a content type classification, and an outline of four to seven sections, which then passes to the first human approval gate.

\subsection{Phase 2: Generation Under Layered Quality Control}

After the outline is approved, the executor generates the full artifact bundle: a hub page built to a fixed layout, four to seven subpages, distribution copy in three hook variations per publishing account, six for the bilingual account, and two direct message templates per account, four for the bilingual account. Generation is constrained by four stacked quality layers. First, exemplar calibration matched by content type: before writing, the executor reads reference guide files of the detected type and matches their depth, formatting density, and tone. Second, source verification rules carried over from Phase 1: no unverified URL, current pricing and version numbers only, and no invented tools or capabilities. Third, a style filter built from 24 documented failure patterns of machine generated prose, covering banned vocabulary, banned rhetorical constructions, and punctuation rules, applied to every sentence. Fourth, self-check lists run on each artifact before it is returned: 4 items for guide content, 8 for distribution copy, including a 200 to 250 word length band, and 4 for message templates. A separate 9 item checklist runs at the level of the whole pipeline before anything reaches the second human gate.

\subsection{Publication, Telemetry, and the Feedback Cycle}

Publication runs through deterministic scripts controlled by the orchestrator, with a cover graphic declared mandatory: a guide cannot be published until its banner is generated and attached. Distribution records carry a keyword unique to each guide, enabling attribution of downstream engagement back to the specific guide that produced it. The autonomous analysis step consumes this telemetry and maintains the strategy file, and generation is bound by two sections of that file: active rules it must follow, and discarded hypotheses it may not retry. Both Phase 0 and Phase 2 are required to read this file before acting, so a measured effect becomes a binding default for the next run rather than a one-time observation. A hook style measured at 1.83 times baseline engagement in the first analysis cycle became the default lead variant this way. The loop also retires its own rules: an earlier default favoring a narrative hook style was withdrawn once a later analysis window measured that class as the worst performing, and the retirement itself is recorded so generation cannot silently revert to it later. These engagement measures function as adoption and demand evidence for topics and framings, not as measures of learning outcome, and the human operator retains final authority over every rule: any rule can be overridden, and the loop adapts to the override in its next cycle rather than resisting it.

\section{Generated Artifacts and Content Design Model}\label{sec:design}

\subsection{Artifact Inventory per Pipeline Run}

A complete run yields two artifact classes: an instructional package and the instrumentation used to acquire an audience for it. The learner facing core is a multi-page guide hosted as a linked page hierarchy. The hub page follows a fixed layout: a community callout, a one sentence description, an author byline, links to every child page, a section listing four to six deliverables, a section listing three to five audience descriptors, a navigation block with a short description and a direct link for each step, and a sources section restricted to evergreen authoritative references. Video sources used during research may not appear in that published sources list. Beneath the hub sit the subpages, one per instructional step, each carrying an icon matched to its step; the specification allows four to seven subpages, and six is the typical count across the deployed corpus, with an observed range of five to seven.

The second artifact class instruments audience acquisition. Each run also produces distribution copy in three hook variations per publishing account, six for the bilingual account, three in English and three in Spanish; two delivery message templates per account, four for the bilingual account, one combining the guide link with a community invitation and one personalized to the specific guide; and a banner specification, consisting of a banner type, a generation prompt, and reference logo queries, rendered through an image generation model conditioned on real logo reference images and attached as the hub cover, a step the workflow marks as mandatory. The published package additionally carries inline screenshots inside the subpages and a branded graphic for the distribution card, produced through the same reference-conditioned image path. Each guide carries an all-caps keyword unique by rule: a prospective reader comments the keyword on the distribution post, an automation detects the comment, and the delivery message returns the guide link, making acquisition volume attributable per guide, with a handful of intentional keyword reuses recorded and discussed in Section~\ref{sec:results}. A derived engagement score, computed as reactions plus 3 times comments plus 2 times shares, is computed per post and treated strictly as adoption and demand evidence informing later topic and hook selection, not as a learning outcome.

The guide-hub core of this inventory was produced for each of the 26 documented runs between March and July 2026. The full multi-account distribution set applied through late June 2026, after which distribution narrowed to a single account at three posts per week.

\subsection{A Three Type Content Taxonomy with Automatic Classification}

Before writing begins, the pipeline classifies the source material into one of three content types using surface signals present in the transcript. Type 1, a step-by-step technical tutorial, is detected from installation steps, configuration, code snippets, terminal commands, API integration, and sequencing language such as first do this, then do that. Type 2, a strategic framework, is detected from methodology, phases, planning, timelines, and growth tactic language. Type 3, a comparison or persuasion guide, is detected from cost analysis, before and after contrasts, return on investment discussion, and argumentation of the form why X is better than Y. When signals are mixed, a dominance rule applies: the material defaults to whichever type covers at least 60 percent of the content. The classification is announced to the human operator together with a planned outline and held for confirmation before drafting starts.

Each type binds to its own template. Type 1 proceeds from an action oriented title and hooks through a what you will learn block, prerequisites, numbered steps that each follow an explanation, a code or configuration block, and an expected result, closing with tips, a cost table, a troubleshooting section, and next steps. Type 2 proceeds from problem framing through a framework overview, phase based steps each carrying goals, tactics, and action items, a key insight section isolating the single mechanism the rest of the guide depends on, and a quick reference summary table. Type 3 proceeds from a thesis through enumerated problems backed by data, a head to head comparison table, a cost analysis with return on investment, an objection handling section, and a final recommendation. Each type also maps to one or two reference exemplars the writer must read before drafting, to calibrate section length, formatting density, tone, and specificity. In deployment the taxonomy skews strongly toward Type 1: roughly 20 of the 26 documented guides are technical tutorials.

\subsection{Embedded Instructional Scaffolding}

Independent of type, every guide embeds a consistent set of scaffolds that parallel conventional instructional design elements. A list of concrete deliverables at the top of the guide functions as a learning objectives analogue. A prerequisites section states required tools, accounts, and subscriptions with explicit cost and time figures; the Type 1 calibration exemplar names a subscription tier tradeoff between a higher tier at 100 dollars a month and a limited tier at 20 dollars a month, the supported operating systems, and a 10 minute time commitment. Worked examples take the form of prompts and configuration blocks the reader can copy directly; the exemplar supplies one master prompt plus four role specific refinement prompts. Each step is followed by feedback on the expected result, including verification commands with an annotated expected output, for example a version check annotated should show v18 or higher, and a short account of what the reader will observe next. Semantic callouts carry fixed meanings throughout: a tip marker for non obvious advice, a warning marker for common mistakes and security concerns, a key insight marker for critical takeaways, and a quick win marker for immediate actions. A troubleshooting section presents problems and solutions in a question and answer format, and guidance drawn from experience is encoded as paired do this and watch out for lists. Finally, a transfer section applies the pattern just taught to new domains; the calibration exemplar closes with five additional use cases of the same multi-agent pattern, three of which quantify the time saved.

Formatting itself is treated as part of the pedagogy and is enforced by checklist: every major heading carries an emoji marker, tables are mandatory for comparisons and pricing, paragraphs are capped at two to four sentences, checkmark and arrow bullets carry distinct meanings, and numbered lists are reserved for sequential steps.

\subsection{The Self-Sufficiency Criterion}

The mastery bar for every guide is stated as a single criterion in the quality checklist: a reader must be able to implement what the guide teaches from the guide alone. This criterion is operationalized through a four-gate checklist run before presentation, covering content, formatting, tone, and lead magnet quality, with concrete supporting rules: all code must be syntactically correct and ready to copy with no unexplained placeholders, every URL must be verified before it is included, and invented tools or capabilities are disallowed. No formal exercises, quizzes, or assessments appear in the current templates; the implementation push at the end of each guide functions as the practical exercise. The absence of a per-guide learning assessment is a limitation noted in Section~\ref{sec:limits}.

\subsection{Produced Examples}

Three deployed guides and the pipeline's own calibration exemplar illustrate the model in practice. The calibration exemplar, Build Your First AI Sales Team in 30 Minutes, is a third-party guide, authored outside the pipeline, that the writer reads before drafting any Type 1 guide; it demonstrates the target register, including explicit position and time budgeting such as setup step 1 of 3, five minutes, no coding experience needed, a cost section, a do and do not section, and the transfer section pattern. How to Build an AI Voice Agent, combining two named tools, is a deployed technical tutorial with six subpages under the keyword VOICE. The Claude Code Cheat Sheet, covering more than 40 tips, commands, and tricks, is a deployed strategic framework with six subpages under the keyword CHEATS, illustrating the punchier phase based structure of Type 2. A fourth case, on Claude Code permission modes, is a deployed technical tutorial whose first subpage carries a five-mode side by side comparison table, showing comparison scaffolding deployed inside a Type 1 guide.

\subsection{Correspondence with the ADDIE Model}

The workflow maps naturally onto the ADDIE instructional design model~\cite{branson1975,spatioti2022}. Analysis corresponds to Phases 0 and 1: the topic scan across configured channels and community forums, cluster scoring with the depth gate, followed by transcript extraction, verification against official documentation, audience pain point mining, and a gap analysis against existing coverage. Design corresponds to the outline artifact, a four to seven section structure where each section carries a title, a description, and a note on what is new, reviewed at the first of the two human approval gates. Development corresponds to Phase 2, where content is written against the type exemplar and passed through the style and self-check filters before the second approval gate. Implementation corresponds to publication: scripted hub and subpage creation, banner attachment, screenshot embedding, and activation of the keyword gated acquisition path. Evaluation corresponds to the feedback cycle: engagement metrics feed the autonomously maintained strategy file that both topic selection and writing consult, and persistent agent memory records approved structures, rejected angles, and recurring review corrections across runs. This evaluation loop measures acquisition demand rather than learning gain, a distinction developed further in Section~\ref{sec:results}.

\section{Production Output and Preliminary Results}\label{sec:results}

\subsection{Content Production at Steady State}

Over a five month span, the pipeline operated in continuous production and published 26 documented guide hubs. This figure is a registry floor: the distribution board additionally records roughly 15 guide slots over the same five months whose hub identifiers were never registered, so the conservative documented count of 26 is used throughout this paper. Each published guide is a multi-page hub, a landing page plus subpages, built to the fixed layout described in Section~\ref{sec:design}. Distribution initially ran three lead magnet slots per week, on Monday, Wednesday, and Friday, mirrored across four publisher accounts for 12 posts per week.

Beyond the hub and its subpages, a single run produces up to 15 social post variations, 10 direct message templates, and a banner specification for the graphic generation step, as detailed in Section~\ref{sec:design}; the full multi-account set applied through late June 2026. The published end state adds a generated banner cover, screenshots embedded in the subpages, a branded distribution graphic, and the coordination board entry from which the team posts. Nothing ships without clearing both the outline gate and the full bundle gate.

\subsection{Operator Effort}

At steady state, the operator reports approximately 1.5 hours of human time per week to take three new guides end to end: research kickoff, both approval gates, publication, and scheduled distribution with the keyword triggered delivery automation configured and live. This figure is an operator-measured self-report rather than an instrumented time study. The surrounding community and distribution workflow adds at most about 1.5 additional hours per week, also operator-reported. Effort concentrates exactly at the points the architecture reserves for judgment: the outline approval gate, the content approval gate, variation selection, post scheduling, and keyword configuration. Research, drafting, source verification, formatting, graphic generation, publishing, and packaging run without operator involvement.

\subsection{The Instrumented Funnel}

Every lead magnet post closes with a fixed two-step call to action: react to the post and comment a guide specific all-caps keyword. The keyword comment is the funnel's intent signal. A third-party keyword detection service monitors post comments and sends an automated direct message containing the guide link. Over the five month window, 51 keyword automations covering the 26 guides registered 3,014 keyword comments and sent 2,803 automated messages; messages from the primary account route recipients to the practice community, while messages from the other accounts link directly to the guide hub. Each hub in turn carries its own community call to action, closing the path from post to guide to membership. Keywords are assigned per guide and are unique by rule, with a handful of documented intentional reuses, one keyword served two related guides in March and May, and four other reuses are recorded in the keyword table, so comment streams approximate per-guide response tracking without additional instrumentation.

The community recorded 612 new members over the five month window and 763 members in total at its close, with 47 percent of new members in the window attributed by the platform to this distribution channel (58 percent of those with a recorded referrer). Both figures come from the platform member export, which lists current members only, so the counts are a floor; both are demand and adoption measures, and their scope is discussed at the end of this section.

\begin{figure}[tp]
\centering
\includegraphics[width=0.97\textwidth]{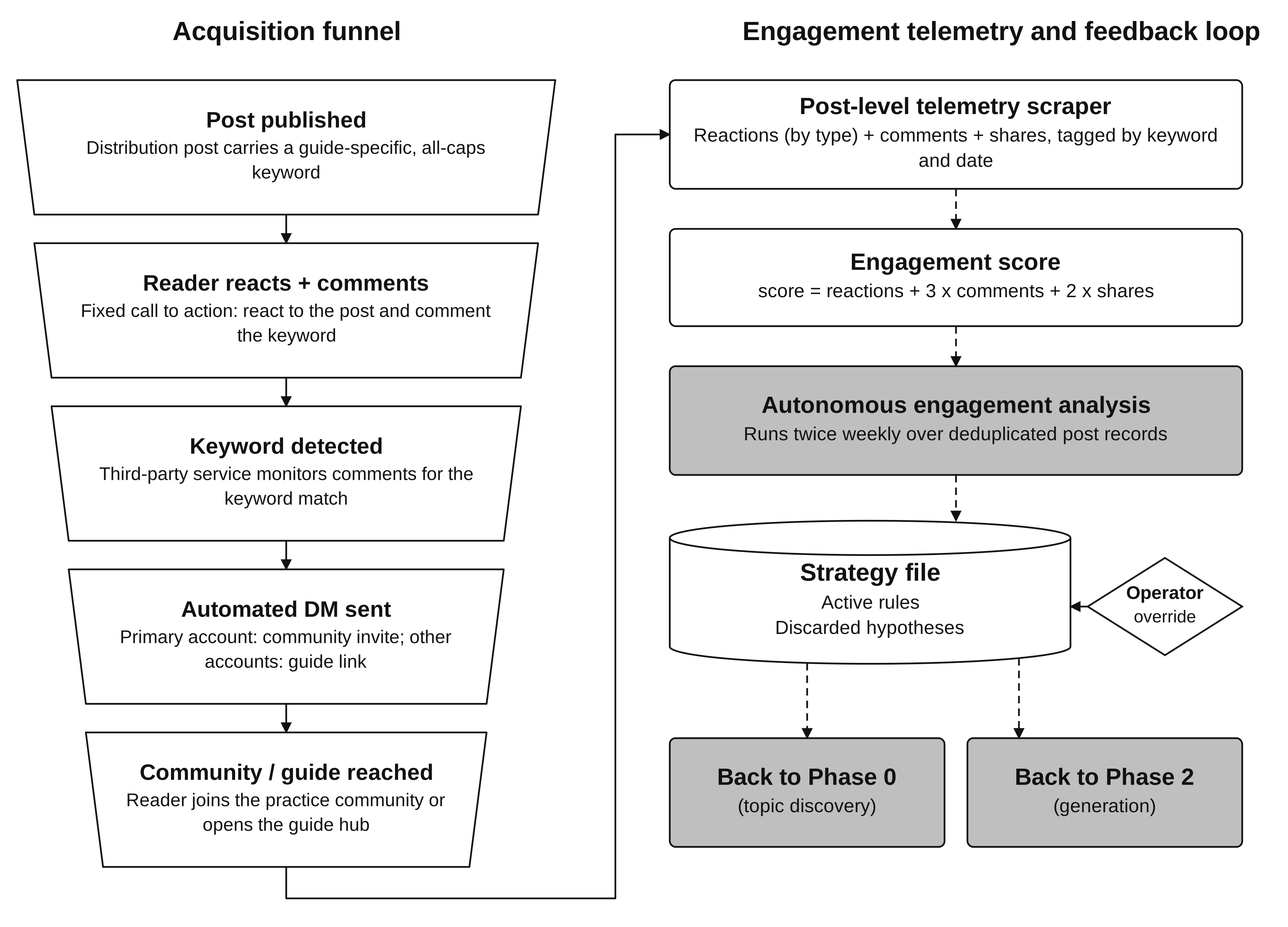}
\caption{The acquisition funnel (left) and the engagement telemetry and feedback loop it feeds (right). A published post carries a keyword; a reader reaction and keyword comment triggers an automated direct message that delivers the guide or a community invitation. Post-level telemetry is scored and analyzed twice weekly, producing active rules and discarded hypotheses in the strategy file, which is read back into Phase 0 and Phase 2 on every subsequent run; the operator can override any rule the loop produces.}
\label{fig:funnel}
\end{figure}

Figure~\ref{fig:funnel} shows the funnel alongside the telemetry and feedback mechanism described in the next two subsections.

\subsection{Engagement Telemetry and Collection Method}

Post-level telemetry is collected by a scraper that records reactions, broken down by reaction type, comments, and shares for each post, tagged with the post's keyword and date. The full posting history is re-collected at analysis time, and records are deduplicated by the pair of post URL and post date; every statistic reported below is computed on these deduplicated records.

The composite metric used throughout is engagement score, defined as reactions plus 3 times comments plus 2 times shares. The top artifact to date is the post for the "Claude Code Skills 2.0" guide: 523 reactions, 1,711 comments, and 7 shares, for a score of 5,670. Two analysis windows have closed so far. The first window, spanning about a month with 16 posts, had a mean score of 447.5 and a median of 349. The second window, spanning roughly two months with 20 posts, had a mean of 300.9 and a median of 263, a 33 percent decline in mean score from the first window. Within the second window, monthly means were 176.8 in the first month with 5 posts, 448.1 in the second month with 8 posts, and 221.3 in the third month with 7 posts. Comments dominate the composite score: across the 63 posts published in the window the correlation between comments and score is 0.999, against 0.928 for reactions, and comments contribute 83 to 93 percent of the total score in the top four posts. This weight toward comments follows directly from the keyword gated design, which converts reader interest into a comment action by construction.

\subsection{Findings from the Feedback Loop}

The engagement analysis step rewrites a strategy file that the writing phase must read before topic selection and drafting, so its findings serve simultaneously as results and as control inputs to later production. Four findings from the deployed loop are reported here.

First, keyword gating is associated with higher engagement: posts carrying a keyword averaged 314.1 across 18 posts, against 182.5 for 2 posts without one. The unequal group sizes make this directional rather than conclusive, but the rule has survived two analysis cycles without being retired.

Second, a contrarian hook style performed at 1.83 times the baseline in the first analysis cycle. In the second window this class remained the strongest, at a mean of 387.5, roughly half its earlier mean of 817, and the loop now pairs it with the hook subject rule below rather than relying on it alone.

Third, the subject of the hook matters. Hooks opening on the reader's own situation averaged 340.5 across 8 posts, hooks opening on a property of the tool averaged 317.0 across 8 posts, and hooks opening on the creator's own activity averaged 189.5 across 4 posts. Every one of the top four posts in the window opened on the reader or the tool. The loop encoded this as an active rule: the first sentence must address the reader or the tool, never the author. The same window retired an earlier default favoring a narrative hook built around the creator's own story, whose class mean had fallen to the worst in the group; the retirement is recorded so generation cannot silently revert to it.

Fourth, topic fatigue is measurable. Topics belonging to a recently repeated family averaged 230.4, against 346.2 for all other topics. The loop also records negative results as discarded hypotheses the writing phase may not retry: pure news announcement posts averaged 154.8, with the single worst post in the window scoring 6.

\subsection{Scope of the Evidence}

The results reported above are adoption and demand measures: production volume, operator effort, engagement, automated delivery, and membership growth. None of them measures learning. The system observes that readers request and receive guides; it does not observe whether they complete them or can later perform the procedures the guides teach. The open confounds and the planned instrument for a follow-up study of learning outcomes are discussed in Section~\ref{sec:limits}. Until then, the claims this deployment supports are limited to production capability, operator effort, and demand.

\section{Discussion}\label{sec:discussion}

The architecture and the production data reported above point to a few design implications beyond the specific system described. The first is that separating deterministic, model bound, and open ended work into distinct node classes is not only an implementation convenience; it changes what evidence is available for evaluation. Because scanning, scoring, and publishing run as fixed scripts rather than agent calls, their behavior is deterministic and auditable across all 26 documented runs, while evaluation attention concentrates on the three phases where an agent actually exercises judgment. A pipeline that routes every step through a single agent call would not offer this same separation, and comparing failure rates across node classes would be far harder to do after the fact.

The second implication concerns grounding. Splitting research into a demand layer and an authority layer answers a specific version of the hallucination problem: a system that mines social platforms for what an audience wants to read is exposed to exactly the unreliable claims those platforms carry. Confining social sources to demand signals, and requiring every factual claim to trace back to an authority source, is a structural rather than a prompted safeguard, in the sense that no instruction to the writing agent, however carefully worded, substitutes for the fact that video sources are excluded from the published sources list by rule. This may explain why the pipeline can absorb topic ideas from informal sources that move quickly without also absorbing their reliability problems, though this paper does not measure factual error rates directly and treats the claim as a design property rather than a validated outcome.

The third implication concerns the feedback cycle itself. The four findings reported in Section~\ref{sec:results} are read back into the pipeline as active rules and discarded hypotheses, so the same telemetry that reports on a guide's reception also changes how the next guide is written. The retirement of the narrative hook default after a later window found it performing worst in its class is a case in point: a static pipeline would have no mechanism for that reversal to happen at all, and the earlier default could persist indefinitely once encoded in a prompt. Whether this adaptive behavior generalizes past the two closed analysis windows reported here, or whether it converges on a small set of rules that then change rarely, is not yet answerable from five months of data and is discussed further as future work in Section~\ref{sec:limits}.

\section{Limitations and Future Work}\label{sec:limits}

Despite the production evidence gathered over five months of deployment, there are some limitations worth mentioning. First, every engagement figure reported in Section~\ref{sec:results} is an adoption and demand measure; none of it observes whether a reader completed a guide or can perform the procedure it teaches, so the pipeline's effect on actual learning outcomes remains an open question rather than a measured one. Second, the operator effort figures are a self-report from the person running the pipeline rather than an instrumented time study, and should be read with that caveat in mind. A later deployment of the same pipeline core (October 2026, a separate product built on the guide-maker modules, six production runs, Sonnet 5.5, list prices) measured 1.07 to 1.61 USD per guide, mean 1.35 USD, of which about 1.00 USD was model inference and 0.35 USD source retrieval, with 4 to 10 minutes of machine time per guide excluding approval wait. Operator time in that deployment was roughly 10 minutes per guide, against about 30 minutes implied by the 1.5 hours per week self-report above. The figures are not directly comparable: the later deployment keeps only the two approval gates in the human loop and leaves keyword gated distribution (variation selection, scheduling and keyword setup) outside it. Both operator figures are estimates, not instrumented time studies. Third, two confounds identified in the tracking file remain unresolved with the current sample: the June engagement spike cannot be separated into topic and timing effects, and the decline in mean engagement between the first and second analysis windows may reflect changes in platform reach, audience fatigue, or topic selection, and the present data cannot distinguish among these explanations.

As future work, interactive assessment modules embedded directly in the guides are planned as the instrument needed for a follow-up study of learning outcomes, since no comparable instrument exists in the current templates. An instrumented time study across a fixed sample of production runs would replace the current self-reported operator effort figures with directly measured ones. Extending the deployment to a second domain that changes just as quickly, outside artificial intelligence tools, would test whether the architecture generalizes past the domain analyzed in this paper. Finally, closing the two open confounds around the June spike and the decline between windows would require collecting topic and timing as separate variables in the telemetry pipeline, rather than inferring them after the fact from aggregate engagement scores.

\section{Conclusion}\label{sec:conclusion}

This paper has presented an agentic pipeline for technology content production, organized as a typed workflow graph and closed by a feedback cycle that rewrites its own generation rules from audience engagement. The architecture separates deterministic scripts, model steps bound to a fixed rubric, open ended agent phases, and human approval gates into four distinct node classes, and splits research into a demand layer and an authority layer so that popularity signals shape topic selection without ever substituting for a verified source. Over five months of continuous deployment the pipeline published 26 documented guides at roughly 1.5 hours of operator time per week, and its feedback cycle produced four measurable engagement findings, one of which reversed an earlier default once later data contradicted it.

The evidence gathered so far establishes production capability, operator effort, and audience demand as adoption measures; it does not establish a learning outcome, and the two are kept explicitly distinct throughout this paper. That distinction sets the direction for the work ahead: an instrumented assessment layer would turn the demand evidence reported here into a measurable claim about what readers actually learn, and a second deployment outside artificial intelligence tools would show whether the typed workflow graph described in this paper is a general architecture for content production in any technical domain that changes as quickly, or a pattern specific to the domain analyzed here.

\begingroup\small
\bibliographystyle{IEEEtran}
\bibliography{refs}
\endgroup

\end{document}